\documentclass[preprint,showpacs,preprintnumbers,amsmath,amssymb,superscriptaddress]{revtex4}

\usepackage{graphicx,amsfonts}
\usepackage{epsfig}
\usepackage{dcolumn}
\usepackage{bm}
\begin{document}

\title{Neutron electric dipole moment in the minimal 3-3-1 model}


\author{G. De Conto}%
\email{georgedc@ift.unesp.br}
\affiliation{
Instituto  de F\'\i sica Te\'orica--Universidade Estadual Paulista \\
R. Dr. Bento Teobaldo Ferraz 271, Barra Funda\\ S\~ao Paulo - SP, 01140-070,
Brazil
}
\author{V. Pleitez}%
\email{vicente@ift.unesp.br}
\affiliation{
Instituto  de F\'\i sica Te\'orica--Universidade Estadual Paulista \\
R. Dr. Bento Teobaldo Ferraz 271, Barra Funda\\ S\~ao Paulo - SP, 01140-070,
Brazil
}

\date{02/09/16}
%
\begin{abstract}
We calculate the electric dipole moment (EDM) for the neutron in the framework of the minimal 3-3-1 model. We assume that the only source of $CP$ violation arises from a complex trilinear coupling constant and two complex vacuum expectation values. However, from the constraint equations obtained from the potential, only one physical phase remains. We find some constraints on the possible values of this phase and masses of the exotic particles.
\end{abstract}


\maketitle




%

\section{Introduction}

The measurement of the electric dipole moment (EDM) of elementary particles is a crucial issue to  particle physics. This is because  for a non-degenerate system, as a nucleus or  an elementary particle, an EDM is possible only if the symmetries under $T$ and $C\!P$ are violated. On one hand, in the Standard Model (SM) framework the only source of $T$ and $C\!P$ violation is the phase $\delta$ in the CKM mixing matrix.
On the other hand, the SM prediction for the neutron electric dipole moment (EDM) is $|d_n|_{\textrm{SM}}\approx10^{-32}\,e\cdot$ cm \cite{Shabalin:1978,
Shabalin:1983,Shabalin:1980,Eeg:1984,Czarnecki:1997}, six orders of magnitude below the actual experimental limit of $d_n\vert_\textrm{exp}=-0.21 \pm 1.82 10^{-26}\,e\cdot \textrm{cm}$ or an upper limit of $3.6\times10^{-26}\,e\cdot \textrm{cm}\,(95\%\,C.L.)$~\cite{Afach:2015sja}.
Hence, we see that in the context of the Standard Model the Kobayashi-Maskawa phase is not enough for explaining an EDM with a value near the experimental limit for both electron and neutron. If the latter case is confirmed in future experiments, it certainly means the discovery of new physics with new $C\!P$ violation sources.

Moreover, cosmology also hints that the SM may not be a complete description and that new $CP$ violating phases must exist in models beyond the SM in order to explain the observed matter--anti-matter asymmetry of the Universe~\cite{Riotto:1999, Morrisey:2012, Dine:2004}. Therefore, we are led to explore alternatives to the SM, in our case we consider the minimal 3-3-1 model (m331 for short) with a heavy sextet~\cite{DeConto:2015eia}.  However, in this work we will only be concerned with the EDM issue. 

The 3-3-1 models are interesting extensions of the standard model which can give some insight in the issue of the number of generations and the value of $\sin^2\theta_W$. Moreover, some of these models having vector-quarks with exotic electric charge and at least one neutral scalar coupling with the exotic quarks can explain the 750 GeV resonance observed at LHC. See for instance \cite{Martinez:2015kmn,Hernandez:2015ywg,Cao:2015scs}, and references therein.
Although this resonance has not been confirmed by recent ATLAS~\cite{ATLAS:2016eeo} and CMS~\cite{CMS:2016crm} data, it is clear that if resonances around 1-2 TeV are discovered in the near future, this model certainly will be able to give it an explanation. 

The EDM of  electron and neutron in the context of the 3-3-1 model with heavy leptons (331HL for short) has been considered in  Refs.~\cite{Montero:1998yw,DeConto:2014fza}.
The representation content in the quark sector is the same as in the minimal 3-3-1 model, but the number of the scalar multiplets in the former are larger and include a scalar sextet, under the $SU(3)_L\otimes U(1)_x$ it transforms as $S\sim(6,0)$, which is needed to give the correct mass to the charged leptons. However, the degrees of freedom of the sextet decouple when its fields  $S$ are heavy, inducing nonrenormalizable interactions between the triplets giving mass to the charged leptons, while the neutrino masses are obtained if we also add right-handed neutrinos transforming trivially under the gauge symmetry of the model and using the type I seesaw mechanism.

The outline of this paper is as follows.
In Sec.~\ref{sec:model} we introduce the representation content of the model. The scalar sector in Subsec.~\ref{subsec:scalars},  while the quarks are presented in Subsec.~\ref{subsec:quarks}. In 
Sec.~\ref{sec:neutronedm}, we calculate the EDM for the neutron. The last section, Sec.~\ref{sec:con}, is devoted to our conclusions. In the Appendices~\ref{sec:matrizesmassa}~--~\ref{sec:verticesgaugescalar} we write all the interactions and mass eigenstates used in our calculations.

\section{The 3-3-1 model}
\label{sec:model}

Here we will work in the framework of the minimal 3-3-1 model with a heavy sextet and right-handed neutrinos studied in Ref.~\cite{DeConto:2015eia}. In this model the electric charge operator is given by
\begin{equation}
\frac{Q}{|e|}=T_{3}-\sqrt{3}T_{8}+X,
\end{equation}
where $e$ is the electron charge, $T_{3,8}=\lambda_{3,8}/2$ (being $\lambda_{3,8}$ the Gell-Mann matrices) and $X$ is the hypercharge operator
associated to the $U(1)_X$ group. In the following subsections we present the field content of the model, with its charges associated to each
group on the parentheses, in the form ($SU(3)_C$, $SU(3)_L$, $U(1)_X$).

\subsection{The scalar sector}
\label{subsec:scalars}

In this, as in other 3-3-1 models, there are many phases in the mixing matrices. Even if the phases in the CKM mixing matrix are absorbed in the quark fields, they appear in the interactions of the fermions with heavy vector and scalar bosons~\cite{Promberger:2007py}. Here we will consider that the only source of $C\!P$ violation phases are in the scalar sector and one of its trilinear couplings. The scalar potential is given in Ref.~\cite{DeConto:2015eia} with a few changes. There, all the VEVs and coupling constants were assumed real. If lepton number is conserved in the scalar potential, there are two linear trilinear interactions with couplings $f_1$ and $f_2$. For the sake of simplicity we will consider only $f_1$ to be complex. Hence, we begin with the following phases (the notation is $x_i=e^{i\theta_i}\vert x_i\vert$): $\theta_\eta,\theta_\rho,\theta_\chi,\theta_{s2}$ and $\theta_1$ of $v_\eta,v_\rho,v_\chi,v_{s2}$ and $f_1$, respectively. The $SU(3)$ transformation $T=\textrm{diag}(e^{-i\theta_\eta}, e^{-i\theta_\rho},e^{i(\theta_\eta+\theta_\rho)})$ allows us to eliminate two phases, $\theta_\eta$ and $\theta_\rho$. After this transformation is done the phase of $v_\chi$ in the trilinear term $f_1\epsilon\,\eta\rho\chi$ becomes $\theta^\prime_\chi=
\theta_\eta+\theta_\rho+\theta_\chi$, and in the trilinear term $f_2(\chi^TS^*\rho+\rho^TS^*\chi)$ the phase of $v_{s2}$ becomes $\theta^\prime_{s2}=\theta_{s2}-\theta_\rho$. Hence we have three phases up to now: $\theta_1,\theta_\chi,\theta_{s2}$ (we have omitted the prime in the phases of $\theta_\chi$ and $\theta_{s2}$). Next the constraint equations involving only the latter phases, that are obtained by
taking the derivatives of the potential with respect to the VEV's, become
\begin{eqnarray}\label{eq:derivadavs}
&& \frac{\partial V }{\partial |v_{s2}|}=\frac{1}{4} \Big[ |v_{s2}|  \left( v_{\chi} ^2 (2  d_1 + d_2 )+2  d_3   v_{\eta} ^2+ v_{\rho} ^2 (2  d_5 + d_6 )+2  |v_{s2}| ^2 (2  e_1 + e_2 )+4  \mu_{s2}^2  \right) \nonumber \\& &+ 2  f_2   v_{\rho}   v_{\chi}  \cos ( \theta_{s2} + \theta_{\chi} )\Big]
\nonumber \\
&& \frac{\partial V }{\partial |v_{\chi}|}= a_3   |v_{\chi}| ^3+\frac{1}{4}  |v_{\chi}|  \left(2  a_4   v_{\eta} ^2+2  a_5   v_{\rho} ^2+ |v_{s2}| ^2 (2  d_1 + d_2 )+4  \mu_{\chi}^2  \right) \nonumber \\& &+\frac{1}{2}  v_{\rho}  \left(\sqrt{2}  |f_1|   v_{\eta}  \cos ( \theta_{f1} + \theta_{\chi} )+ f_2   |v_{s2}|  \cos ( \theta_{s2} + \theta_{\chi} )\right),
\nonumber \\ && 
\frac{\partial V }{\partial v_{\eta}}=\frac{1}{2} \left( v_{\eta}  \left(2  a_1   v_{\eta} ^2+ a_4   |v_\chi| ^2+ a_6   v_{\rho} ^2+ d_3   |v_{s2}| ^2+2  \mu_{\eta}^2  \right)+\sqrt{2}  |f_1|   v_{\rho}   |v_\chi|  \cos ( \theta_{f1} + \theta_{\chi} )\right),\nonumber \\ &&
\frac{\partial V }{\partial v_{\rho}}=\frac{1}{4} \Big[ v_{\rho}  \left(4  a_2   v_{\rho} ^2+2  a_5   |v_\chi| ^2+2  a_6   v_{\eta} ^2+ |v_{s2}| ^2 (2  d_5 + d_6 )+4  \mu_{\rho}^2  \right) \nonumber \\& &+2 \sqrt{2}  |f_1|     |v_\chi|  v_{\eta} \cos ( \theta_{f1} + \theta_{\chi} )+2  f_2   |v_{s2}|   |v_\chi|  \cos ( \theta_{s2} + \theta_{\chi} )\Big],
\nonumber \\ && 
\frac{\partial V }{\partial \theta_{s2}}=-\frac{1}{2} f_2 v_\rho |v_{s2}| |v_\chi| \sin (\theta_{s2}+\theta_{\chi}),
\nonumber \\ && 
\frac{\partial V }{\partial \theta_{\chi}}=-\frac{1}{2}  v_{\rho}   |v_\chi|  \left(\sqrt{2}  |f_1|   v_{\eta}  \sin ( \theta_{f1} + \theta_{\chi} )+ f_2   |v_{s2}|  \sin ( \theta_{s2} + \theta_{\chi} )\right),
\end{eqnarray}
where $\theta_{s2}$, $\theta_\chi$ and  $\theta_{f_1}$ are the complex phases for the VEVs and the coupling constant. 
At the potential minimum all derivatives above should be zero, in doing so, from $\partial V /\partial \theta_{s2}=\partial V /\partial \theta_{\chi}=0$ we find $\theta_{s2}=\theta_{f_1}=-\theta_\chi$, and using this in the other constraints we obtain
\begin{eqnarray}
&& \mu_{s2}^2  = -\frac{ |v_{s2}|   |v_\chi| ^2 (2  d_1 + d_2 )+2  d_3   v_{\eta} ^2  |v_{s2}| + v_{\rho} ^2  |v_{s2}|  (2  d_5 + d_6 )+2  |v_{s2}| ^3 (2  e_1 + e_2 )+2  f_2   v_{\rho}   |v_\chi| }{4  |v_{s2}| },\nonumber \\ && 
 \mu_{\chi}^2  = -\frac{ |v_\chi|  \left(4  a_3   |v_\chi| ^2+2  a_4   v_{\eta} ^2+2  a_5   v_{\rho} ^2+ |v_{s2}| ^2 (2  d_1 + d_2 )\right)+2 \sqrt{2}  |f_1|   v_{\eta}   v_{\rho} +2  f_2   v_{\rho}   |v_{s2}| }{4  |v_\chi| },
\nonumber \\ &&
 \mu_{\eta}^2  = -\frac{2  a_1   v_{\eta} ^3+ a_4   v_{\eta}   |v_\chi| ^2+ a_6   v_{\eta}   v_{\rho} ^2+ d_3   v_{\eta}   |v_{s2}| ^2+\sqrt{2}  |f_1|   v_{\rho}   |v_\chi| }{2  v_{\eta} },
\nonumber \\ &&
 \mu_{\rho}^2  = -\frac{4  a_2   v_{\rho} ^3+2  |v_\chi|  \left( a_5   v_{\rho}   |v_\chi| +\sqrt{2}  |f_1|   v_{\eta} \right)+2  a_6   v_{\eta} ^2  v_{\rho} + v_{\rho}   |v_{s2}| ^2 (2  d_5 + d_6 )+2  f_2   |v_{s2}|   |v_\chi| }{4  v_{\rho} }.
\end{eqnarray}

This scalar potential leads to mass matrices where the analytical solutions for the mass eigenstates are not available. Therefore, in the same vein as in \cite{DeConto:2015eia}, we will work with approximate mass matrices, where we assume that $|v_\chi|/|v_{s2}|>>1$ and also disregard some non-diagonal elements assuming that the diagonal elements dominate. These mass matrices and their corresponding eigenstates and eigenvectors are presented in Appendix~\ref{sec:matrizesmassa}. 

\subsection{Quarks}
\label{subsec:quarks}

In the quark sector there are two anti-triplets and one triplet, all left-handed, besides the corresponding right-handed singlets:
\begin{equation}
Q_{mL}=\left(\begin{array}{c}
d_{m} \\ -u_{m} \\ j_{m}
\end{array}\right)_L \sim \left(3,3^{*},-1/3\right) ,\qquad
Q_{3L}=\left(\begin{array}{c}
u_{3} \\ d_{3} \\ J
\end{array}\right)_L \sim \left(3,3,2/3\right)
\end{equation}

\begin{equation}
u_{\alpha R} \sim \left(3,1,2/3\right) ,\quad d_{\alpha R} \sim \left(3,1,-1/3\right) ,\quad j_{mR} \sim \left(3,1,-4/3\right),
\quad J_{R} \sim \left(3,1,5/3\right)
\end{equation}
where $m=1,2$ and $\alpha=1,2,3$. The $j_m$ exotic quarks have electric charge -4/3 and the $J$ exotic quark has electric charge 5/3 in units of $\vert e \vert$.

The Yukawa interactions between quarks and scalars are given by:
\begin{eqnarray} 
-\mathcal{L}^q_Y &=& \bar{Q}_{mL} \left[ G_{m\alpha} U^{'}_{\alpha R} \rho^*+\tilde{G}_{m\alpha}D^{'}_{\alpha R} \eta^* \right]+
\bar{Q}_{3L} \left[ F_{3\alpha}U^{'}_{\alpha R} \eta + \tilde{F}_{3\alpha}D^{'}_{\alpha R} \rho \right] \nonumber \\&  +&
\bar{Q}_{mL}G'_{mi}j_{iR}\chi^* + \bar{Q}_{3L}g_J J_R \chi +H.c.,
\label{q1}
\end{eqnarray}
where we omitted the sum in $m$, $\alpha$ and $i=1,2,3$, $U^{'}_{\alpha R}=(u^\prime\,c^\prime\,t^\prime)_R$ and $D^{'}_{\alpha R}~=~
(d^\prime\,s^\prime\,b^\prime)_R$. $G_{m\alpha}$, $\tilde{G}_{m\alpha}$, $F_{3\alpha}$, $\tilde{F}_{3\alpha}$, $G'_{mi}$ and $g_J$ are the coupling constants.

From Eq.~(\ref{q1}), we obtain that the exotic quarks have the following interactions with the charged scalars
\begin{eqnarray}
-\mathcal{L}_j&=&
\bar{\tilde{j}}_L[\mathcal{O}^u
V^U _R\, U_R +\mathcal{O}^dV^D_R\, D_R] +\frac{\sqrt2}{\vert v_\chi\vert}\,\bar{D}_LV^D_L \left(\begin{array}{ccc}
m_{j_1}\chi^+ & 0 & 0\\
0 & m_{j_2}\chi^+&0\\
0&0&m_J\chi^{--}
\end{array}\right)\tilde{j}_R
\nonumber \\&+&
\frac{\sqrt2}{\vert v_\chi\vert}\,\bar{U}_L \,V^U_L \left(\begin{array}{ccc}
m_{j_1}\chi^{++}& 0 & 0\\
0 & m_{j_2}\chi^{++}&0\\
0&0&m_J\chi^{-}
\end{array}\right)\tilde{j}_R+
H.c.
\label{jjJ}
\end{eqnarray}
with $\tilde{j}=(j_1\,j_2\,J)^T$, $U_{L,R}=(u\,c\,t)^T_{L,R}$ and $D_{L,R}=(d\,s\,b)^T_{L,R}$ denoting the mass eigenstates.
We have defined the matrices 
\begin{equation}
\mathcal{O}^u=\left(\begin{array}{ccc}
G_{11}\rho^{--}&G_{12}\rho^{--}&G_{13}\rho^{--}\\
G_{21}\rho^{--}&G_{22}\rho^{--}&G_{23}\rho^{--}\\
F_{31}\eta^+_2 &F_{32}\eta^+_2 &F_{33}\eta^+_2 \\
\end{array}\right),\;\;\mathcal{O}^d=
\left(\begin{array}{ccc}
\tilde{G}_{11}\eta^-_2&\tilde{G}_{12}\eta^-_2 &\tilde{G}_{13}\eta^-_2 \\
\tilde{G}_{21}\eta^-_2 &\tilde{G}_{22}\eta^-_2 &\tilde{G}_{23}\eta^-_2 \\
\tilde{F}_{31}\rho^{++}  &\tilde{F}_{32}\rho^{++}  &\tilde{F}_{33}\rho^{++}  \\
\end{array}\right).
\label{J}
\end{equation}

In Eq.~(\ref{jjJ}) we have assumed that the mass matrix in the $j_1,j_2$ sector is diagonal, i.e.,  $G'_{12}=G'_{21}=0$. In this case $G_{ii}=\vert G_{ii}\vert e^{i\theta_\chi}$ and $g_J=\vert g_J\vert e^{i\theta_\chi}$. After absorbing the $\theta_\chi$ phase in the masses we have $\vert g_J\vert=m_J\sqrt{2}/\vert v_\chi\vert$ and $\vert G_{ii}\vert=m_{j_i}\sqrt{2}/\vert v_\chi\vert$.  
We have also used the fact that if $U^{'}_{L,R}$ and $D^{'}_{L,R}$ denote the symmetry eigenstates and $U_{L,R}$ and $D_{L,R}$ the mass eigenstates. They are related by unitary matrices as follows: $U^{'}_{L,R}=\left( V_{L,R}^U\right)^\dagger U_{L,R}$ and $D^{'}_{L,R}=\left( V_{L,R}^D\right)^\dagger D_{L,R}$ in such a way that
$V_{L}^{U} M^u V_{R}^{U\dagger}=\hat{M}^u=diag(m_u,m_c,m_t)$ and
$V_{L}^{D} M^d V_{R}^{D\dagger}=\hat{M}^d=diag(m_d,m_s,m_b)$.

In terms of the mass eigenstates we can write the Yukawa interactions in Eqs.~(\ref{jjJ}) and (\ref{J}) as in Appendix~\ref{sec:quarksscalars}, where the charged scalars have already been projected on the physical $Y^-_2,Y^{--}$. In this appendix we wrote only the interactions which appear in the EDM diagrams. 

Using as input the observed quark masses and the mixing matrix in the quark sector, $V_{CKM}=V_L^U
V_L^{D\dagger}$~\cite{Agashe:2014kda}, the numerical values of the matrices $V^{U,D}_{L,R}$ were found to be \cite{Machado:2013jca}:
\begin{eqnarray}
&& V^U_L=\left(\begin{array}{ccc}
	-0.00032 & 0.00433 & 0.99999 \\
	0.07163 & -0.99742 & 0.00434 \\
	-0.99743 & -0.07163 & -0.00001 \\
\end{array}\right),\nonumber \\&&
 V^D_L\!\!=\!\!\left(\begin{array}{ccc}
	0.004175 & -0.209965 & 0.97761 \\
	0.03341 & -0.977145 & -0.209995 \\
	-0.999525 & -0.03052 & -0.004165 \\
\end{array}\right)
\label{vudl331},
\end{eqnarray}

In the same way we obtain the $V^{U,D}_R$ matrices:
\begin{eqnarray}
&& V^U_R=\left(\begin{array}{ccc}
	-0.4544 & 0.13857 & 0.87996 \\
	0.82278 & -0.31329 & 0.47421 \\
	-0.34139 & -0.93949 & -0.02834 \\
\end{array}\right),\nonumber  \\&&
 V^D_R\!\!=\!\!\left(\begin{array}{ccc}
	-0.0001815 & -0.325355 & 0.94559 \\
	0.005976 & -0.945575 & -0.325345 \\
	-0.999982 & -0.00559 & -0.002115 \\
\end{array}\right).
\label{vudr331}
\end{eqnarray}

It should be noted that the product $V^U_L V^{D\dagger}_L$ of the matrices above correspond to the CKM matrix when the modulus is considered. The known quark masses depend on both $v_\eta$ and $v_\rho$. The values of the matrices $V^{U,D}_{L,R}$ were obtained by using $v_\rho=54$ GeV and $v_\eta=240$. For the reasons implying the values for the VEVs see Ref.~\cite{Dias:2006ns} . The matrices given in Eqs.~(\ref{vudl331}) and (\ref{vudr331}) give the correct quark masses (at the $Z$-pole given in Ref.~\cite{Machado:2013jca}) and the CKM matrix if the Yukawa couplings are:  $G_{11}=1.08,G_{12}=2.97,G_{13}=0.09,G_{21}=0.0681,
G_{22}=0.2169,G_{23}=0.1\times10^{-2}$,
$F_{31}=9\times10^{-6},F_{32}=6\times10^{-6},F_{33}=1.2\times10^{-5}$, $\tilde{G}_{11}=0.0119,
\tilde{G}_{12}=6\times10^{-5},\tilde{G}_{13}=2.3\times10^{-5},\tilde{G}_{21}=(3.2 - 6.62)\times10^{-4},\tilde{G}_{22}=
2.13\times10^{-4},\tilde{G}_{23}=7\times10^{-5}$, $\tilde{F}_{31}=2.2\times10^{-4},
\tilde{F}_{32}=1.95\times10^{-4},\tilde{F}_{33}=1.312\times10^{-4}$. All these couplings should be multiplied by $\sqrt{2}$, it is a conversion factor from the notation used in \cite{Machado:2013jca} to our notation. We also took the central values of the matrices $V^D_{L,R}$ presented in this reference for our calculations.

The numerical solutions in (\ref{vudl331}) and (\ref{vudr331}) are not unique and difficult to be obtained, and we cannot claim that we are exploring all the parameter space. However, they are sufficiently realistic for considering that the results obtained using them are also realistic and a possibility for the constraints of the mass of the particles in the model. 

\section{The neutron EDM}
\label{sec:neutronedm}

In the framework of quantum field theory (QFT) the EDM of a fermion is described by an effective Lagrangian
\begin{equation}
\mathcal{L}_{EDM}=-i\sum_{f}\frac{d}{2}\bar{f}\sigma^{\mu\nu}\gamma_{5}fF_{\mu\nu}
\label{eq1}
\end{equation}
where $d$ is the magnitude of the EDM, $f$ is the fermion wave function and $F_{\mu\nu}$ is the electromagnetic tensor. This Lagrangian
gives rise to the vertex
\begin{equation}
\label{vertice_MDE}
\Gamma^{\mu}=id\sigma^{\mu\nu}q_{\nu}\gamma_{5}
\end{equation}
where $q_{\nu}$ is the photon's momentum.

Since the EDM is an electromagnetic property of a particle, its Lagrangian depends on the interaction between the particle and the electromagnetic field. To find the EDM one must consider all the diagrams for a vertex between the particle and a photon. The sum of the
amplitudes will be proportional to
\begin{equation}
\label{vertice_generico_MDM_eletron}
\Gamma^{\mu}\left(q\right)=F_{1}\left(q^{2}\right)\gamma^{\mu}+\cdots+F_{3}\left(q^{2}\right)\sigma^{\mu\nu}\gamma_{5}q_{\nu}
\end{equation}
Comparing with Eq.~(\ref{vertice_MDE}), we can see that $d=\textrm{Im}[F_3(0)]$.  

\begin{figure}
\includegraphics[width=0.75\textwidth]{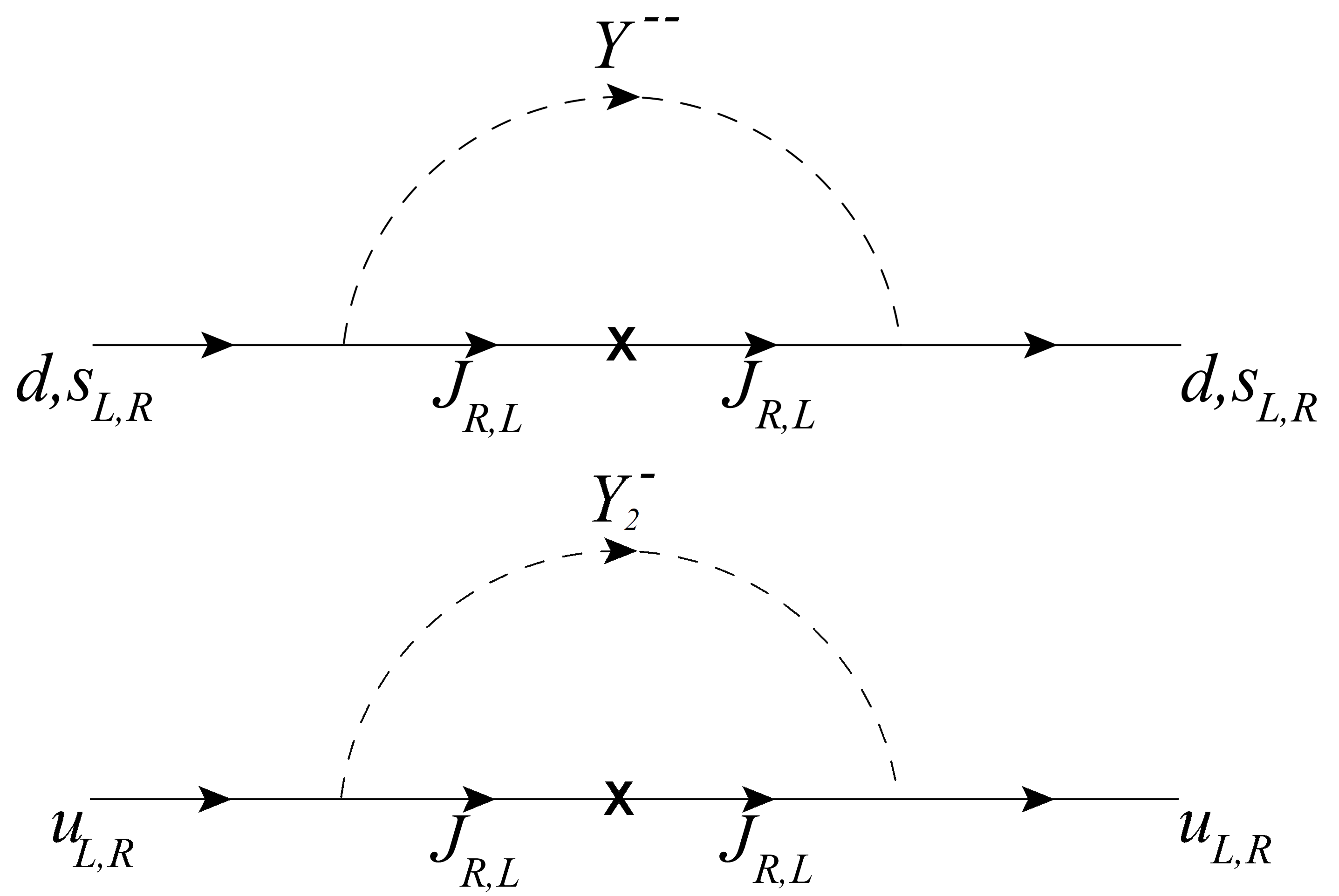}
\caption{Diagrams contributing to the neutron EDM. For each diagram in the figure should be considered the case where the photon line is connected to the scalar line and the case where it is connected to the fermion line. All the left-right combinations should be considered as well.}
\label{fig:edm1}
\end{figure}

We will assume here that the only source of $CP$ violation is the phase $\theta_\chi$ as found out in Sec.~\ref{subsec:scalars}. Considering the diagrams given in Fig.~\ref{fig:edm1} we find an expression for the neutron EDM in this 3-3-1 model. For each diagram we calculate the contribution to the EDM given by each quark, with the total EDM of the neutron, assuming contributions of the quarks $u,d,s$, is written by
\begin{equation} \label{eq:mde_331}
d_n\vert_Y=d_u g^u_T+ d_dg^d_T+ d_sg^s_T 
\end{equation}
where in the quark model $g^u_T=4/3$, $g^d_T=-1/3$. Here we use the form factors obtained from lattice QCD:  $g^d_T=0.774$, $g^u_T=-0.233$ and $g^s_T=0.008$~\cite{Bhattacharya:2015esa,Bhattacharya:2015wna,Chien:2015xha}.

The analytical expressions for each quark contribution are:
\begin{eqnarray}
\left.\frac{d_d}{e\cdot\textrm{cm}}\right\vert_Y&=&\left\lbrace \textrm{Im}\Big[(K_{J_LD_R})_{31}(K_{D_LJ_R})_{13}\Big]-\textrm{Im}\Big[(K^\dagger_{D_LJ_R})_{31}(K^\dagger_{J_LD_R})_{13}\Big]\right\rbrace 
\nonumber \\ &\cdot& \Big[I^{dJY}_1 + I^{dJY}_2) \Big],
\nonumber \\& =&
-(197 \times 10^{-16}\;\text{GeV})\left[   \frac{2\sqrt{2}|v_\rho|}{|v_\rho|^2+|v_\chi|^2}(V^D_L)_{13} \sum_{k} (V^D_R)_{1k} \tilde{F}_{3k} \right]\nonumber \\ &\cdot& \Big[I^{dJY}_1+ I^{dJY}_2\Big]\,\sin(2\theta_\chi) m_J,
\label{edmn1}
\end{eqnarray}
where $Y$ denotes $Y^{++}$. 

\begin{eqnarray}
\left.\frac{d_s}{e\cdot\textrm{cm}}\right\vert_Y&=&\left\lbrace
\textrm{Im}\Big[(K_{J_LD_R})_{32}(K_{D_LJ_R})_{23}\Big]-\textrm{Im}\Big[(K^\dagger_{D_LJ_R})_{32}(K^\dagger_{J_LD_R})_{23}\Big]\right\rbrace
\nonumber \\ &\cdot& \Big[I^{dJY}_1 + I^{dJY}_2) \Big],
\nonumber \\& =&
-(197 \times 10^{-16}\;\text{GeV})\left[ 
\frac{2\sqrt{2}|v_\rho|}{|v_\rho|^2+|v_\chi|^2}(V^D_L)_{23} \sum_{k}
(V^D_R)_{2k} \tilde{F}_{3k} \right]\nonumber \\ &\cdot&
\Big[I^{dJY}_1+ I^{dJY}_2\Big]\,\sin(2\theta_\chi) m_J,
\label{edmn2}
\end{eqnarray}


Similarly, considering the figures involving the $u$ quark in Fig.~\ref{fig:edm1}, in this case the charged scalar is $Y^+_2$,
\begin{eqnarray}
\left.\frac{d_u}{e\cdot\textrm{cm}}\right\vert_{Y_2}&=&\left\lbrace \textrm{Im}\Big[(K_{J_LU_R})_{31}(K_{U_LJ_R})_{13}\Big]-\textrm{Im}\Big[(K^\dagger_{U_LJ_R})_{31}(K^\dagger_{J_LU_R})_{13}\Big]\right\rbrace \Big[I^{uJY_2}_1 + I^{uJY_2}_2) \Big]
\nonumber \\& =&
-(197 \times 10^{-16}\;\text{GeV})\left[  \frac{2\sqrt{2}|v_\eta|}{|v_\rho|^2+|v_\chi|^2}(V^U_L)_{13} \sum_{k} (V^U_R)_{1k} F_{3k} \right]\nonumber \\ &  \cdot & \Big[I^{uJY_2}_1+ I^{uJY_2}_2\Big]\,\sin(2\theta_\chi) m_J,
\label{edmn3}
\end{eqnarray}
and the integrals $I^{dJY}_{1,2}$ are given by

 \begin{equation} \label{eq:I1}
 I^{qQY}_1\equiv I_1(m_q,m_J,m_Y)=-\frac{m_JQ_J}{32\pi^2}\int\limits_{0}^{1} dz
 \frac{1+z}{(m^2_J-zm^2_q) (1-z)+m_Y^2 z},
 \end{equation}
 and
 \begin{equation} \label{eq:I2}
 I^{qQY}_2\equiv  I_2(m_q,m_J,m_Y)=-\frac{m_JQ_Y}{32 \pi^2}\int\limits_{0}^{1} dz
 \frac{z}{[m^2_J-(1-z)m^2_q] z+m_Y^2(1-z)},
 \end{equation}
 where $m_{Q}$ and $m_q$  denote the masses of the exotic quark and the known quarks, respectively. Also, $m_Y$ is the mass of the scalar in the diagram, $Y^{++},Y^+_2$. Finally, $Q_Y$ and $Q_J$ denote the electric charge of the scalar and the quark in the loop, respectively (1 for the $Y_2^+$, 2 for the $Y^{++}$ and 5/3 for the $J$ quark). In all the calculations above, we have used the interactions in Sec.~\ref{sec:quarksscalars}. 

On the equations above we have considered the $V^U_{L,R}$ and $V^D_{L,R}$ matrices to be real, that is because we considered the numerical results presented in~\cite{Machado:2013jca} for such matrices and for the Yukawa couplings [see Eqs.~(\ref{vudl331}) and (\ref{vudr331})].

\begin{figure}[tbp]
\centering 
\includegraphics[width=.75\textwidth,origin=c,angle=0]{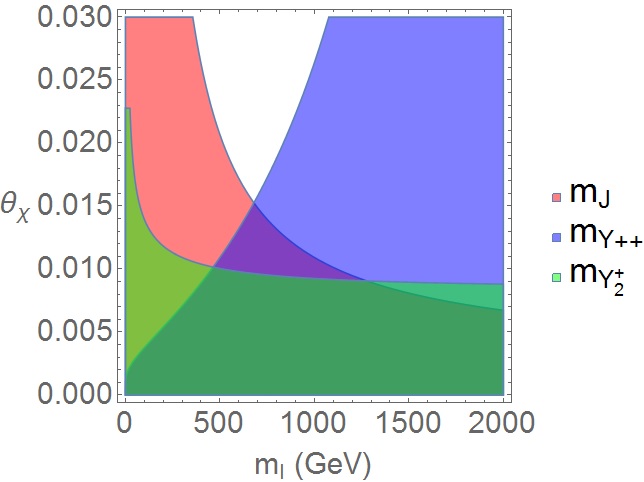}
\hfill
\caption{
Allowed values for the exotic particle masses and $\theta_\chi$. The shaded regions indicates the values for $m_J$ (red), $m_{Y^{++}}$ (blue), $m_{Y_2^+}$ (green), and $\theta_\chi$ on which the neutron EDM is within 1$\sigma$ of the the experimental results. On the horizontal axis, $m_I$ indicates the mass of the particle according to the color of the shaded region, while the value for the complex phase $\theta_\chi$ is in the vertical axis (in radians). The values for the other parameters of the model are fixed (see text for more information).}
\label{fig:EDM_neutron}
\end{figure}

Using eq. (\ref{eq:mde_331}) and considering that it respects the actual experimental limit \cite{Afach:2015sja} ($ d_n\vert_Y =-0.21 \pm 1.82 \times 10^{-26}\,e\cdot \textrm{cm}$) we obtain the graph in Fig. \ref{fig:EDM_neutron}. The shaded regions indicates the allowed values for the exotic particle mass and the VEV's complex phase, in which the neutron EDM is within 1$\sigma$ of the experimental results. For each plot we fixed the parameters as: $|v_{\chi}|=$ 2000, $m_J=$ 1000, $m_{Y_2^+}=$ 300, and $m_{Y^{++}}=$ 500 (all in GeV); these are fixed when such parameters are not varied on each analysis. We can see in Fig. \ref{fig:EDM_neutron} that lower values for $m_J$ allows greater freedom in the possible values of $\theta_\chi$ (red region), while the opposite happens for $m_{Y^{++}}$. As for $m_{Y_2^+}$, low values allow a larger parameter range for $\theta_\chi$, but for values above 1000 GeV the complex phase cannot be higher than 0.010 radians, since the upper limit of the green region becomes nearly horizontal.

We can understand the above results from the analytical expressions in (\ref{edmn1}), (\ref{edmn2}), and (\ref{edmn3}), that all contributions to the nEDM are proportional to $\sin(2\theta_\chi) m_J$, hence large $m_J$ implies small phases $\theta_\chi$ (or other values where the sine is small), while lower values for this mass allows a greater range of values for the complex phase, as can be sees in the figure. For the scalar masses, the analysis is more intricate. The nEDM dependence on these parameters can be seen in Eqs.~(\ref{eq:I1}) and (\ref{eq:I2}), where the masses appear on the denominator of the integrands, the higher the scalar mass, the smaller the integrand value. The $d$ and $s$ quarks give a positive contribution with their integrals depending on the $Y^{++}$ mass, while the $u$ quark gives a negative contribution with its integral depending on the $Y_2^+$ mass. Therefore, high values for $m_{Y^{++}}$ and small values for $m_{Y^+_2}$ imply a small nEDM, setting the complex phase aside. In this case, the complex phase respects the experimental limit in a broader range of values. The opposite happens when $m_{Y^{++}}$ is small and $m_{Y^+_2}$ large. More important, we cannot forget that there is a balance between each quark contribution, where the smallest possible EDM is when the negative contribution from the $u$ quark cancels the positive contributions from the $d$ and $s$ quarks, allowing any value for the complex phase.

Experimental limits on the masses of the exotic particles are very model dependent, however here we assumed values compatible with experimental searches. For the exotic quark $J$ with electric charge 5/3 we considered $M_J>840$~\cite{Chatrchyan:2013wfa,Aad:2015mba}. 
For the singly charged scalar masses two lower limits can be considered $M_{Y^+_2}>300(800)$~\cite{ATLAS:2016qiq}, and for the doubly charged scalars 
we use $m^2_{Y^{++}}>580$~\cite{ATLAS:2016pbt}. Note that in our plots we start from null masses in the horizontal axis. In this manner it is possible to have a better understanding of the numerical results from a mathematical perspective.

\section{Conclusions}
\label{sec:con}

In the framework of the 3-3-1 models, the neutron EDM was calculated in Refs.~\cite{Montero:1998yw}. However, at that time we knew nothing about the unitary matrices in the quark sector, $V^{U,D}_{L,R}$. Notwithstanding, after the results from Ref.~\cite{Machado:2013jca} it is possible to make more realistic calculations of the EDM once now the number of free parameters is lower than before. In fact, once the values of $ v_\rho$ and $v_\eta$ are obtained, the quark masses and the CKM matrix determine, not necessarily unequivocally, the unitary matrices in the quark sector. At this level, the unknown parameters are the phase $\theta_\chi$, the masses of the exotic quarks and scalars, and  the orthogonal matrix which diagonalize the mass matrix of the $C\!P$ even neutral scalars. 

Here we have shown that the neutron EDM imposes a constraint in the new mechanism of $C\!P$ violation arising from the complex phase in the $\chi$ triplet VEV.
Moreover, from the EDM of the neutron at 1-loop order we were able to set limits on the masses of the exotic quarks and the complex phase of $v_\chi$, which are compatible with the search of these sort of fields at the LHC and Tevatron~\cite{Davey:2014tka}. It seems that in this model we have a situation similar to that in supersymmetric theories, in which the EDM's are larger than the SM prediction and are appropriately suppressed only by the phases. This is the so called SUSY $C\!P$-problem. See Ref.~\cite{Pospelov:2005pr,Ritz:2009zz} and references therein. However, we stress again that we have considered only the soft $C\!P$ violation present in the model. In fact, it has other $C\!P$ hard violating sources. Beside the phase $\delta$ in the CKM matrix, the matrices $V^{U,D}_{L,R}$ are also complex with, in principle, six arbitrary phases. It is possible that three of the phases in $V^D_L$ can be absorbed in the exotic quarks $J,j_1$ and $j_2$, but there is no more freedom to absorb the phases in $V^U_L$, and we also have the phases in $V^{U,D}_R$. Notwithstanding, these phases will appear in the vertices shown in Appendix.~\ref{sec:quarksscalars}. 

There are also contributions from the chromo-electric dipole moment (CEDM), mainly that of the top quark~\cite{Chien:2015xha}. This effect is important to higher order calculations and may further constrain our values for $\theta_\chi$. However, this goes beyond the scope of our work and we hope these issues will be considered elsewhere.

\acknowledgements

G. De Conto would like to thank CNPq for the financial support and V. P. would like to thank CNPq for partial support. We would also like to thank Dr. Jordy de Vries on his remarks about the neutron form factors and the CEDM.

\newpage
\appendix

\section{Scalar mass eigenstates and eigenvalues}
\label{sec:matrizesmassa}

Although the m331 has a rich scalar sector including a sextet~\cite{Foot:1992rh}, in the context of the model with a heavy sextet\cite{DeConto:2015eia} the model seems like the 3-3-1 model with heavy leptons~\cite{Pleitez:1992xh} in which only three triplets are needed for breaking the gauge symmetry and give mass to all charged fermions. However, the degrees of freedom of the scalar sextet still exist but, in the approximation used here, they do not mix with the other scalars of the same charge.
As we said in Sec.~\ref{sec:model}, so we are considering the case when $\vert v_\chi\vert/|v_{s2}|\gg1$ and disregarding some off-diagonal elements, we are able to find the following mass matrices for the scalar sector. All mass matrices, but that of the real neutral scalar, are block diagonal in the approximation used here.

\begin{itemize}
\item Singly charged scalars 1, in the basis $(\rho^+, \eta_1^+)(M_1^+)^2(\rho^-, \eta_1^-)$
	
\begin{equation}
(M_1^+)^2=\frac{1}{2} \left(
\begin{array}{cc}
-\frac{2 |v_\chi|  \left(\sqrt{2}  |f_1|   v_\eta \right)-2  a_9   v_\eta ^2 v_\rho }{4 v_\rho } & \frac{ |f_1|   |v_\chi| }{\sqrt{2}}-\frac{ a_9   v_\eta  v_\rho }{2}  \\
\frac{ |f_1|   |v_\chi| }{\sqrt{2}}-\frac{ a_9   v_\eta  v_\rho }{2} & \frac{1}{2} \left( a_9   v_\rho ^2-\frac{\sqrt{2}  |f_1|   v_\rho  |v_\chi| }{v_\eta }\right)  \\
\end{array}
\right),
\label{aae1}
\end{equation}
$h^+$ does not mix, it is already a mass eigenstates with
\begin{equation}
M^2_{h^+_1}=\frac{ d_4   v_\eta ^2}{4}-\frac{1}{4} v_\rho  \left( d_6   v_\rho +\frac{2  f_2   |v_\chi| }{|v_{s2}| }\right)
\label{aae2}
\end{equation}
	
\item Singly charged scalars 2, in the basis $(\chi^+, \eta_2^+)(M_2^+)^2(\chi^-, \eta_2^-)$ 
\begin{equation}
(M_2^+)^2=\frac{1}{2} \left(
\begin{array}{cc}
-\frac{\sqrt{2}  |f_1|   v_\eta  v_\rho - a_7   v_\eta ^2 |v_\chi| }{2 |v_\chi| } & \frac{1}{2} e^{i  \theta_\chi  } \left( a_7   v_\eta  |v_\chi| -\sqrt{2}  |f_1|   v_\rho \right)  \\
\frac{1}{2} e^{-i  \theta_\chi  } \left( a_7   v_\eta  |v_\chi| -\sqrt{2}  |f_1|   v_\rho \right) & \frac{2}{4} |v_\chi|  \left( a_7   |v_\chi| -\frac{\sqrt{2}  |f_1|   v_\rho }{v_\eta }\right) \\
\end{array}
\right),
\label{aae3}
\end{equation}
$h^+_2$ does not mix, and 
\begin{equation}
M^2_{h^+_2}=\frac{ d_4   v_\eta ^2}{4}-\frac{1}{4} |v_\chi|  \left( d_2   |v_\chi| +\frac{2  f_2   v_\rho }{|v_{s2}| }\right)
\label{aae4}
\end{equation}

\item Doubly charged scalars, in the basis $(\chi^{++}, \rho^{++})(M^{++})^2(\chi^{--}, \rho^{--})$
	
\begin{equation}
(M^{++})^2=\frac{1}{2} \left(
\begin{array}{cc}
\frac{v_\rho  \left( a_8   v_\rho  |v_\chi| -\sqrt{2}  |f_1|   v_\eta \right)}{2 |v_\chi| } & \frac{1}{2} e^{i  \theta_\chi  } \left( a_8   v_\rho  |v_\chi| -\sqrt{2}  |f_1|   v_\eta \right) \\
\frac{1}{2} e^{-i  \theta_\chi  } \left( a_8   v_\rho  |v_\chi| -\sqrt{2}  |f_1|   v_\eta \right) & \frac{|v_\chi|  \left( a_8   v_\rho  |v_\chi| -\sqrt{2}  |f_1|   v_\eta \right)}{2 v_\rho }  
\end{array}
\right),
\label{aae5}
\end{equation}
\end{itemize}
where $H^{++}_{1,2}$ are already mass eigenstates with masses
\begin{eqnarray}
&& M^2_{H^{++}_1}= \frac{ d_6   v_\rho ^2}{4}-\frac{1}{4} |v_\chi|  \left( d_2   |v_\chi| +\frac{2  f_2   v_\rho }{|v_{s2}| }\right),\nonumber \\&&
M^2_{H^{++}_2}=\frac{1}{4} |v_\chi|  \left( d_2   |v_\chi| -\frac{2  f_2   v_\rho }{|v_{s2}| }\right)-\frac{ d_6   v_\rho ^2}{4}.
\label{aae6}
\end{eqnarray}

With the mass matrices above we are able to find the following mass eigenstates for the scalar sector. From the matrix in (\ref{aae1}) we obtain the eigenvectors 
\begin{itemize}
\item Singly charged scalars 1 ($h^-$ does not mix)
\begin{equation}
\left(
\begin{array}{c}
\rho^- \\ \eta_1^- 
\end{array}
\right)=
\frac{1}{\sqrt{1+\frac{v_\eta^2}{v_\rho^2}}}
\left(
\begin{array}{cc}
1 & - \frac{v_\eta}{v_\rho}  \\
\frac{v_\eta}{v_\rho} & 1  \\
\end{array}
\right)
\left(
\begin{array}{c}
G_1^- \\ Y_1^- 
\end{array}
\right)
\label{aae7}
\end{equation}
where the respective eigenvalues are given by	
\begin{eqnarray}
&& m_{G^-_1}=0,\nonumber \\ &&
m^2_{Y^-_1}=\frac{\left(v_\eta ^2+v_\rho ^2\right) \left( a_9   v_\eta  v_\rho -\sqrt{2}  |f_1|   |v_\chi| \right)}{4 v_\eta  v_\rho }.
\label{sse8}
\end{eqnarray}
	
\item Singly charged scalars 2 ($h^-$ does not mix). From (\ref{aae3}) we obtain
\begin{equation}
\left(
\begin{array}{c}
\chi^- \\ \eta_2^- 
\end{array}
\right)=
\frac{1}{\sqrt{1+\frac{v_\eta^2}{|v_\chi|^2}}}
\left(
\begin{array}{cc}
1 & e^{i\theta_\chi} \frac{v_\eta}{|v_\chi|}  \\
- e^{-i\theta_\chi} \frac{v_\eta}{|v_\chi|} & 1  \\
\end{array}
\right)
\left(
\begin{array}{c}
G_2^- \\ Y_2^- 
\end{array}
\right)
\label{aae9}
\end{equation}
and the eigenvalues are
\begin{eqnarray}
&& m_{G^-_2}=0,\nonumber \\ &&
m^2_{Y^-_2}=\frac{\left(v_\eta ^2+|v_\chi| ^2\right) \left( a_7   v_\eta  |v_\chi| -\sqrt{2}  |f_1|   v_\rho \right)}{4 v_\eta  |v_\chi| }.
\label{aae10}
\end{eqnarray}
	
\item Doubly charged scalars ($H^{--}_1$ and $H^{--}_2$ do not mix). From (\ref{aae5}) we obtain 
	
\begin{equation}
\left(
\begin{array}{c}
\chi^{--} \\ \rho^{--} 
\end{array}
\right)=
\frac{1}{\sqrt{1+\frac{v_\rho^2}{|v_\chi|^2}}}
\left(
\begin{array}{cc}
1 & e^{i \theta_\chi} \frac{v_\rho}{|v_\chi|}  \\
-e^{-i \theta_\chi} \frac{v_\rho}{|v_\chi|} & 1 \\
\end{array}
\right)
\left(
\begin{array}{c}
G^-- \\ Y^{--} 
\end{array}
\right)
\label{aae11}
\end{equation}
with the eigenvalues:	
\begin{eqnarray}
&& m_{G^{--}}=0,\nonumber \\ && 
m^2_{Y^{--}}=\frac{\left(v_\rho ^2+|v_\chi| ^2\right) \left( a_8   v_\rho  |v_\chi| -\sqrt{2}  |f_1|   v_\eta \right)}{4 v_\rho  |v_\chi| }.
\label{aae12}
\end{eqnarray}
	
\end{itemize}

\section{Quark-scalar interactions}
\label{sec:quarksscalars}

From Eqs.~(\ref{jjJ}) and (\ref{J}) we obtain the Yukawa interactions with the charged scalars that contribute to the neutron EDM. 
Interactions among $D_L$-type and $J_R$ quarks:
\begin{equation}
-\mathcal{L}_{YD_LJ_R}=\bar{D}_L
K_{D_LJ_R}\,J_R  Y^{--},
\label{ec1}
\end{equation}
where $J_R=(0\,0\,J)_R$ and with
\begin{equation}
K_{D_LJ_R}=\frac{\sqrt{2}e^{2i \theta_\chi}}{\vert v_\chi\vert\sqrt{1+\frac{|v_\chi|^2}{|v_\rho|^2}}}\,V^{D}_L
\left(\begin{array}{ccc}
0 & 0 & 0 \\ 0 & 0 & 0 \\ 0 & 0 & m_J
\end{array}\right).
\end{equation}

Interactions among $J_L$ and $D_R$-type quarks:
\begin{equation}
-\mathcal{L}_{YJ_LD_R}=\bar{J}_L
K_{J_LD_R}\,D_RY^{++},
\label{ec2}
\end{equation}
with
\begin{equation}
K_{J_LD_R}=\frac{|v_\chi|}{\sqrt{|v_\rho|^2+|v_\chi|^2}}\,\left( \begin{array}{ccc}
0 & 0 & 0 \\ 0 & 0 & 0 \\ \tilde{F}_{31} & \tilde{F}_{32} & \tilde{F}_{33}
\end{array}\right)V_R^{D\dagger}.
\end{equation}

Interactions among $J_L$ and $U_R$-type quarks:

\begin{equation}
	-\mathcal{L}_{YJ_LU_R}=\bar{J}_L 
	K_{J_LU_R}\,U_R Y^+_2,
	\label{ec3}
\end{equation}
with
\begin{equation}
K_{J_LU_R}= \frac{|v_\chi|}{\sqrt{|v_\chi|^2+|v_\eta^2|}}
\left( \begin{array}{ccc}
0 & 0 & 0 \\ 0 & 0 & 0 \\ F_{31} & F_{32} & F_{33}
\end{array}\right)V^{U\dagger}_R
\end{equation}

Interactions among $U_L$-type and $J_R$-type quarks:

\begin{equation}
-\mathcal{L}_{YU_LJ_R}=\bar{U}_L 
K_{U_LJ_R}\,J_R Y^-_2,
\label{ec4}
\end{equation}
with
\begin{equation}
K_{U_LJ_R}= \frac{\sqrt{2}|v_\eta|e^{2i \theta_\chi}}{|v_\chi|\sqrt{|v_\chi|^2+|v_\eta^2|}}
V^{U}_L
\left(\begin{array}{ccc}
0 & 0 & 0 \\ 0 & 0 & 0 \\ 0 & 0 & m_J
\end{array}\right)
\end{equation}

For the numerical values for the matrices $V^{U,D}_{L,R}$ see Eq.~(\ref{vudl331}) and (\ref{vudr331}) and for those of the parameters in Eqs.~(\ref{ec1}) - (\ref{ec4}) see below Eq.~(\ref{vudr331}).
Notice that both matrices left- and right-handed survive in different interactions in the scalar sector.

\section{Scalar-photon interactions}
\label{sec:verticesgaugescalar}

Now, from the covariant derivatives of the scalar's lagrangian
\begin{equation}
\mathcal{L}_S=\sum_{i=\eta,\rho,\chi} (D^i\phi_i)^\dagger(D^i\phi_i)
\label{dc}
\end{equation}
where $D^i$ are the covariant derivatives, 
 we can find the vertexes for the interactions between scalars and photons. The $A_\mu Y^+_{1,2}Y^-_{1,2}$ vertexes are both equal to $ie(k^- -k^+)_\mu$, and the vertex $A_\mu Y^{++}Y^{--}$ is $2ie(k^- -k^+)_\mu $. The terms $k^{+}$ and $k^-$ indicate, respectively, the momenta of the positive and negative charge scalars. The momenta are all going into the vertex and the modulus of the electric charge of the electron is given by
\begin{equation}
e=g\frac{t}{\sqrt{1+4t^2}}=g\sin\theta_W
\end{equation}
with $t=s_W/\sqrt{1-4s^2_W}$.

\newpage

\newpage

\end{document}